\documentclass[aps,prb,twocolumn]{revtex4}

 \usepackage{graphicx}

\usepackage{amsmath}

\newcommand{\EqLabel}[1] { \label{#1} }
\newcommand{\mb}[1] {\vec{#1}}

\begin{document}
\title{ The effect of the Abrikosov vortex phase on spin and charge states in
magnetic semiconductor-superconductor hybrids}

\author{Tatiana G. Rappoport} \affiliation{Instituto de F\'{\i}sica,
Universidade Federal do Rio de Janeiro, Caixa Postal 68.528-970, Rio de Janeiro, Brazil}

\author{Mona Berciu} \affiliation{Department of Physics and Astronomy,
University of British Columbia, Vancouver, BC V6T 1Z1, Canada}

\author{Boldizs\'{a}r Jank\'{o}} \affiliation{Department of Physics,
University of Notre Dame, Notre Dame, Indiana 46556}
\affiliation{Materials Science Division, Argonne National Laboratory,
Argonne, Illinois 60439}

\date{\today}

\begin{abstract}
We explore the possibility of using the inhomogeneous magnetic field
carried by an Abrikosov vortex in a type-II superconductor to
localize spin-polarized textures in a nearby magnetic semiconductor
quantum well. We show how Zeeman-induced localization induced by a
single vortex is indeed possible, and use these results to
investigate the effect of a periodic vortex array on the transport
properties of the magnetic semiconductor. In particular, we find an
unconventional Integer Quantum Hall regime, and predict directly
testable experimental consequences due to the presence of the
periodic spin polarized structure induced by the superconducting
vortex lattice in the magnetic semiconductor.
\end{abstract} 

\maketitle

\section{Introduction}

The possibility of exploring both charge and spin degrees of freedom
in carriers has attracted the attention of the condensed matter
community in recent years. The discovery of ferromagnetism in
diluted magnetic semiconductors (DMS) like GaMnAs~\cite{ohno96},
opened the opportunity to explore precisely such independent spin
and charge manipulations. A fundamental property of diluted magnetic
semiconductors (DMS) (both new III-Mn-V and the more established
II-Mn-VI systems) is that a relatively small external magnetic field
can cause enormous Zeeman splittings of the electronic energy
levels, even when the material is in the paramagnetic
state~\cite{furdyna88,lee00}. This feature can be used in
spintronics applications as it allows separating states with
different spin. For instance, Fiederling {\it et al.} had
successfully used a II-Mn-VI DMS under effect of low magnetic fields
in spin-injection experiments ~\cite{fiederling99}.

Another utilization of this feature has been discussed by
us~\cite{berciu03,redlinski05a,redlinski05b,berciu05}: Due to the
giant Zeeman splitting, a magnetic field with considerable spatial
variation can be a very effective confining agent for spin polarized
carriers in these DMS systems.

Producing a non-uniform magnetic field with nanoscale spatial
variation inside DMS systems can be an experimental challenge. One
option is the use of nanomagnets deposited on the top of a DMS
layer. In fact, nanomagnets have already been used as a source of
non-homogeneous magnetic field \cite{peeters93,freire00}. Freire
{\it et al.}~\cite{freire00}, for example, have analyzed the case of
a normal semiconductor in the vicinity of nanomagnets. In such case,
the non-homogeneous magnetic field modifies the exciton
kinetic-energy operator and can weakly confine excitons in the
semiconductor. In contrast, we have examined a DMS in the vicinity
of nanomagnets with a variety of
shapes~\cite{berciu03,redlinski05a,redlinski05b}. In these cases, the
confinement is due the Zeeman interaction which is hundreds of times
stronger than the variation of kinetic-energy in DMS.

Another possibility for obtaining the inhomogeneous magnetic fields is the use of superconductors. Nanoscale field singularities appear naturally in
the vortex phase of superconducting (SC) films. Above the lower
critical field B$_{c_1}$, in the Abrikosov vortex phase,
superconducting vortices populate the bulk of the sample, forming a
flux lattice, each vortex carrying a quantum of magnetic flux
$\phi_0/2=h/(2e)$. The field of a single vortex is non-uniformly
distributed around a core of radius $r\sim \xi$ (where $\xi$ is the
coherence length) decaying away from its maximum value at the vortex
center over a length scale $\lambda$ (where $\lambda$ is the
penetration depth).

A regular  two dimensional electron gas (2DEG) in the vicinity of
superconductors in a vortex phase has already been the subject of both
theoretical~\cite{rammer87,brey93,chang94,nielsen95,gumbs95, oh99,
matulis00, wang04} and
experimental studies~\cite{bending90,geim92,danckwerts00}. In this
context, because the Land\'{e} $g$ factor (and thus the Zeeman
interaction) is very small in normal
semiconductors, the main consequence of the inhomogeneous magnetic fields is a change
in the kinetic-energy term of the Hamiltonian. The flux tubes do not
produce bound states, but act
basically as scattering centers~\cite{bending90,nielsen95}.  The
Zeeman interaction was either
neglected or treated as a small perturbation whose main consequence
was to broaden the already known Hofstadter butterfly subbands. 

In the
work presented here, we consider the case where the 2DEG is confined inside a
DMS. Due to the giant Zeeman interaction in paramagnetic DMSs (whose
origin is briefly explained in 
Section II), the Zeeman 
interaction becomes the dominant term in these systems, leading to
appearance of  {\it bound states} inside the flux tubes. These 
 play a
central role in our work and qualitatively change the nature of the
magnetotransport in these systems, compared to the ones previously studied.

\begin{figure}[t]

\includegraphics[width=0.85\columnwidth]{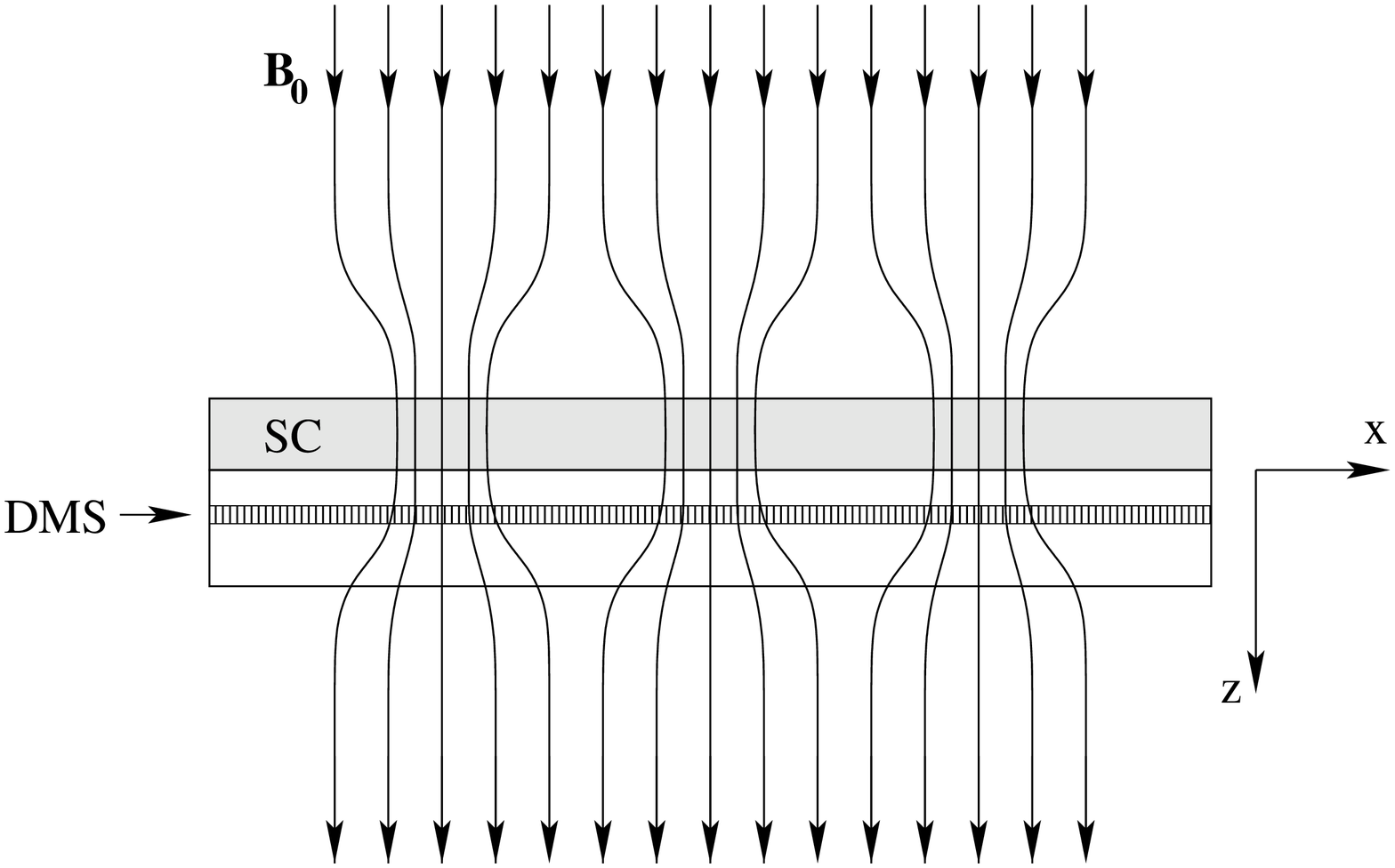}
\caption{Sketch of the type-II SC - DMS heterostructure in a uniform external
magnetic field.} \label{Fig1}

\end{figure}

In this article we study a SC film deposited on top of a diluted
magnetic semiconductor quantum well under the influence of an
external magnetic field $\vec{B}_0 = B_0 \vec{e}_z$ (see
Fig. \ref{Fig1}), for all values of $B_0$ between the two possible
asymptotic limits. For very low applied fields, the SC film is populated with
few isolated vortices, each vortex producing a highly inhomogeneous
magnetic field which localizes spin polarized states in the DMS.
In section \ref{isolated} we obtain numerically the energy
spectrum and the wave functions for the bound states localized by
the vortex field of an isolated vortex in the DMS QW.

However, the main advantage of using SC vortices to generate
confining potentials in the DMSs is the possibility of varying the
distance between them by adjusting the external magnetic field.  For
increasing values of the external magnetic field $B_0$, the vortices
in the SC are organized in a triangular lattice. The lattice spacing
$a$ is related to the
applied magnetic field  by $B_0=\phi_0/(\sqrt{3}a^2)$. In this
limit, the triangular vortex lattice in the SC leads to a periodically
modulated magnetic field inside the neighboring DMS layer.

Since the magnetic field produced by the SC flux lattice creates an
effective spin dependent confinement potential for the carriers in the
DMS, the properties of the superconductors will be reflected in the
energy spectrum of the DMS. In particular, as the lattice spacing and
spatial dependence of the non-uniform magnetic field (our confining
potentials) change with $B_0$, we have
a peculiar system in the DMS where both the lattice spacing and the depth
of the potentials are modified by an applied magnetic field.

For low external magnetic fields, the overlap between the magnetic
fields of independent vortices is small and as a consequence, the
giant Zeeman effect produces deep effective potentials. The trapped
states of the isolated vortex discussed in section \ref{isolated}
widen into energy bands of spin polarized states.  The width of the
bands is defined by the exponentially small hopping $t$ between
neighboring trapped states. However, since the flux through each
unit cell is $\phi/\phi_0=q/p=1/2$, the energy bands will be those
of a triangular Hofstadter butterfly~\cite{claro79}. As long as the
hopping $t$ is small compared to the spacing between consecutive
trapped states, this Hofstadter problem corresponds to the regime of
a dominant periodic modulation, which can be treated within a simple
tight-binding model \cite{hofstadter76,claro79} and each band is
expected to split into $p$ (in this case, $p=2$) magnetic subbands.
This band-structure has unique signatures in the magnetotransport, as
discussed below. Its measurement would provide a clear
signature of the Hofstadter butterfly, which is currently a matter
of considerable experimental
interest~\cite{schlosser96,albrecht01,melinte04}.

On the other hand, for a high applied external magnetic field $B_0$, the
magnetic fields of different vortices begin to overlap
significantly. In this limit, the total magnetic field in the DMS
layer is almost homogeneous. Its average is $B_0\vec{e}_z$, but it has
a small additional periodic modulation (the same effect can be
achieved at a fixed $B_0$ by increasing the distance $z$ between the
SC and DMS layers). For such quasi-homogeneous magnetic fields, the
Zeeman interaction can no longer induce trapping; instead it reverts
to its traditional role of lifting the spin degeneracy. The small
periodic modulation insures the fact that the system still corresponds
to a $\phi/\phi_0 =1/2$ Hofstadter butterfly, but now in the other
asymptotic limit, namely that of a weak periodic modulation. In this
case, each Landau level is expected to split into $q$ subbands, with a
bandwidth controlled by the amplitude of the weak modulation. The case
$\phi/\phi_0=1/2$ has $q=1$ and there are no additional gaps in the
bandstructure. As a result, one expects to see the usual IQHE in
magnetotransport in this regime. To summarize, as long as there is a
vortex lattice, the setup corresponds to a $\phi/\phi_0=1/2$
Hofstadter butterfly irrespective of the value of external magnetic
field $B_0$.  Instead, $B_0$ controls the amplitude of the periodic
modulation, from being the large energy scale (small $B_0$) to being a
small perturbation (large $B_0$).

In section \ref{2d}, we use a unified theoretical approach to analyze
how the 2-D modulated magnetic field produced by the vortex lattice
affects the free carries in the DMS QW, and the resulting
bandstructures. We obtain the energy spectrum 
 going from the asymptotic limit of a very small periodic
modulation to the asymptotic limit of isolated vortices. In the later
case, we are able to reproduce the results obtained in the Section
III. 

As one of the most direct signatures of the Hofstadter butterfly, in Section
\ref{IQHE} we discuss the fingerprints of the band structures
obtained in section \ref{2d} on the magnetotransport properties of
these 2DEG, in particular their Hall (transversal) conductance. This
is shown to change 
significantly as one tunes the external magnetic field  between the
two asymptotic limits. Finally, in Section VI we 
discuss the significance of these results.

\section{Giant Zeeman effect in paramagnetic Diluted Magnetic Semiconductors}

One of the most remarkable properties of the DMS is the giant Zeeman
effect they exhibit in their paramagnetic state, with effective
Land\'e factors of the charge carriers on the order of 10$^2$ -
10$^3$. Since this effect plays a key role in determining the
phenomenology we analyze in this work, we briefly review its origin in
this section, using a simple mean-field picture.

The exchange
interaction of an electron with the impurity spins $\vec{S}_i$ located
at positions $\mathbf{R}_i$, is:
\begin{equation}
H_{ex}= \sum_i J(\vec{r}-\vec{R}_i)\vec{S}_i\cdot\vec{s}.
\end{equation}
where $J(\vec{r})$ is the exchange interaction. Within a mean-field
approximation (justified since each carrier interacts with many
impurity spins) $\vec{S}_i\cdot \vec{s}\rightarrow
\langle\vec{S}_i\rangle \vec{s}+\vec{S}_i \langle \vec{s}\rangle$, and
the average exchange energy felt by the electron becomes:
\begin{eqnarray} \label{eq2}
E_{ex}&=& N_0 x \alpha \langle \vec{S}
\rangle \vec{s}
\end{eqnarray}
Here, $x$ is the molar fraction of Mn dopants, $N_0$ is the number of
unit cells per unit volume, $\langle \vec{S} \rangle$ is the average
expectation value of the Mn spins, and $\alpha= \int_{}^{}d\vec{r}
u^*_{c,\vec{k}=0}(\mb{r}) J(\mb{r})u_{c,\vec{k}=0}(\mb{r})$ is an
integral over one unit cell, with $u_{c,\vec{k}=0}(\mb{r})$ being the
periodic part of the conduction band Bloch wave-functions. The usual
virtual crystal approximation, which averages over all possible
location of Mn impurities, has been used. Exchange fields for holes
can be found similarly; they are somewhat different due to the
different valence-band wave-functions.

In a ferromagnetic DMS, this
terms explains the appearance of a finite magnetization below $T_C$:
as the Mn spins begin to polarize, $\langle\vec{S}\rangle\ne 0$, this
exchange interaction induces a polarization of the charge carrier
spins $\langle \mb{s} \rangle \ne 0$. In turn, exchange terms like
$\vec{S}_i \langle\vec{s}\rangle$ further polarize the Mn, until
self-consistency is reached.

In paramagnetic DMS, however,
$\langle\vec{S}\rangle= 0$ and charge carrier states are
spin-degenerate. The spin-degeneracy can be lifted if an external
magnetic field $\mb{B}$ is applied. Of course, the usual Zeeman
interaction $-g \mu_B \vec{s}\cdot \vec{B}$ is present, although this
is typically very small. Much more important is the indirect coupling
of the charge carrier spin to the external magnetic field, mediated by
the Mn spins. The origin of this is the exchange energy of
Eq. (\ref{eq2}), and the fact that in an external magnetic field, the
impurity spins acquire a finite polarization $\langle \vec{S} \rangle
\parallel \vec{B}$, $|\langle\vec{S}\rangle| = S {\cal B}_S (g\mu_B S
B/(k_BT))$, where $S$ is the value of the impurity spins ($5/2$ for
Mn) and ${\cal B}_S$ is the Brillouin function (for simplicity of
notation, we assume the same bare $g$-factor for both charge carriers
and Mn spins; also, here we neglect the supplementary contribution
coming from the $\vec{S}_i \langle\vec{s}\rangle$ terms, since
typically an impurity spin interacts with few charge carriers). The
total spin-dependent interaction of the charge-carriers is, then:


\begin{equation} \EqLabel{eq3}
{\cal H}_{ex}= -g \mu_B \mb{s}\cdot \mb{B}+ N_0 x \alpha \langle \vec{S}
\rangle \vec{s} = g_{\rm eff} \mu_B
\mb{s}\cdot \mb{B}
\end{equation}
where
\begin{equation}
g_{\rm eff}= -g+\frac{N_0 x \alpha}{\mu_B B} S {\cal B}_S
\left(\frac{g\mu_B S B}{k_BT}\right)
\end{equation}
for electrons, with an equivalent expression for holes. For low
magnetic fields, $g_{\rm eff}$ becomes independent of the value of
$B$, although it is a function of $T$ and $x$. Its large effective
value is primarily due to the strong coupling $\alpha$ (large
$J(\mb{r})$) between charge carriers and Mn spins.

In the calculations shown here, we use as DMS parameters an effective
charge-carrier mass $m=0.5 m_e$ and $g_{\rm eff} = 500$, unless
otherwise specified. Such values are reasonable for holes in several
DMS, such as GaMnAs and CdMnTe.

\section{Isolated vortex\label{isolated}}

In this section we discuss the case where the free carriers in a
narrow DMS quantum well are subjected to the magnetic field created by
a single SC vortex. This situation is relevant for very low applied
fields, when the density of SC vortices is very low.

We first need to
know the magnetic field induced by the SC vortex in the DMS QW. For an
isotropic superconductor, this problem was solved by Pearl
\cite{pearl64}. The field outside of the SC is the free space solution
matching the appropriate boundary conditions at the SC surface.  Let
$r=\sqrt{x^2+y^2}$ be the {\em radial} distance measured from the
vortex center, while $z$ is the distance away from the edge of the
superconductor (here, $z$ is the distance between the SC film and the
DMS QW, see Fig.~\ref{Fig1}).  The radial and transversal components
of the magnetic field created by a single vortex in the DMS layer are
\cite{pearl64,kirtley99,carneiro00,brandt95}:
\begin{eqnarray}
B^{(v)}_r(r,z)=\frac{\phi_0}{4\pi\lambda^2}\int_0^{\infty}{k dk
\frac{J_1(kr)\exp(-kz -\frac{1}{2}\xi^2k^2)}{\tau (k+\tau)}}
&&\label{br}\\
B^{(v)}_z(r,z)=\frac{\phi_0}{4\pi\lambda^2}\int_0^{\infty}{k dk
\frac{J_0(kr)\exp(-kz- \frac{1}{2}\xi^2k^2)}{\tau (k+\tau)}}
&&\label{bz}
\end{eqnarray}
$\phi_0=h/e$ is the quantum of magnetic flux, $\lambda$ and $\xi$ are
the penetration depth and the correlation length of the
superconductor, $\tau = \sqrt{k^2+\lambda^{-2}}$ and $J_{\nu}(\zeta)$
are Bessel functions. The term $e^{-\frac{1}{2}\xi^2 k^2}$ is a
cut-off introduced in order to account for the effects of the finite
vortex core size, which are not included in the London
theory\cite{brandt95}.

\begin{figure}[t]
\includegraphics[width=0.9\columnwidth]{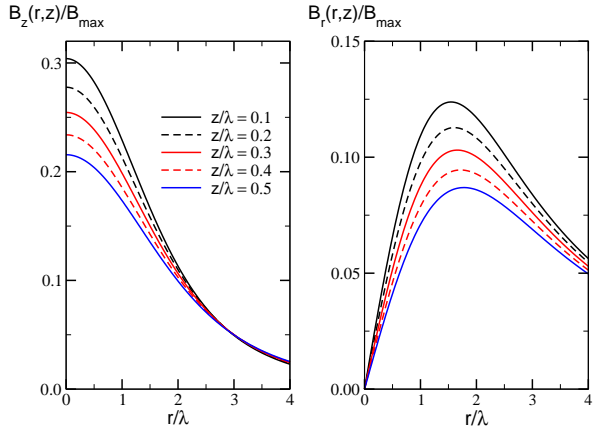}
\caption{The transversal (left) and radial (right) components of the
magnetic field created by a single vortex in Nb ($\xi = 35$ nm,
$\lambda = 40$ nm), at different distances $z$ (in units of
$\lambda$).  $B_{\rm max}={\phi_0}/{(4\pi\lambda^2)}\sim 0.207$ T. }
\label{Fig2ab}
\end{figure}

\begin{figure}[b]
\includegraphics[width=0.9\columnwidth]{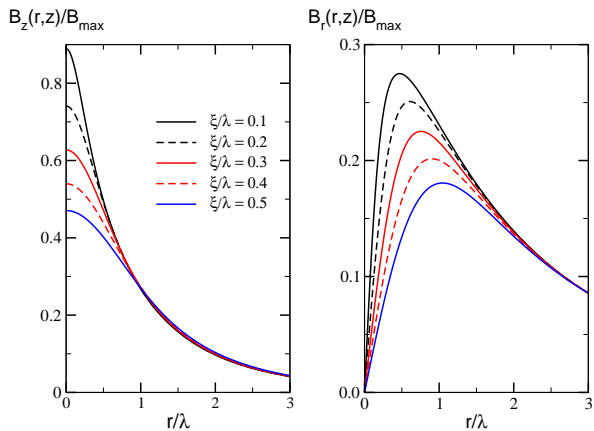}
\caption{The transversal (left) and radial (right) components of the
magnetic field created by a single vortex at a distance
z=0.1$\lambda$, for different values of the coherence length (in units
of $\lambda$. Here, $\lambda = 40$ nm and $B_{\rm
max}={\phi_0}/{(4\pi\lambda^2)}\sim 0.207$ T. }
\label{Fig2a}
\end{figure}

In Figs. \ref{Fig2ab} and \ref{Fig2a} we show typical magnetic field
distributions for different ratios of $\xi/\lambda$ and of
$z/\lambda$. As expected, the transversal component is largest under
the vortex core, and decreases with increasing $z$ (see
Fig. \ref{Fig2ab}). The radial component is zero for $r=0$, has a
maximum for $r \sim \xi$ and then decays fast.  This magnetic field is
qualitatively similar to that created by cylindrical nanomagnets
\cite{berciu03}. The field distribution depends significantly on the
properties of the SC. Its maximum value is bounded by
$B_{max}={\phi_0}/{(4\pi\lambda^2)}$.  For a fixed $\lambda$, higher
values of $B$ and larger gradients (which are desirable for our
problem) occur for smaller ratios $\xi/\lambda$ (see
Fig. \ref{Fig2a}). It follows that the ideal SC candidates would be
extreme type-II (with $\xi/\lambda \ll 1$), and also have small
penetration depths $\lambda$. In order to avoid unnecessary
complications due to pinning, the superconductor should also have low
intrinsic pinning ({\em e.g.}, NbSe$_2$~\cite{bhattacharya93} or
MgB$_2$~\cite{bugoslavsky01}).  Incidentally, NbSe$_2$ is also attractive since
it was shown that it can be deposited via MBE onto GaAs
\cite{yamamoto94}. In the calculations shown here, we use SC parameters
characteristic of Nb: $\lambda =40 \ {\rm nm}$, $\xi = 35 {\rm nm}$.

We now analyze the effects of this magnetic field on the DMS charge
carriers. Using a parabolic band approximation, the effective
Hamiltonian of a charge carrier inside the DMS QW, in the presence of
the magnetic field $\vec{B}^{(v)}(r,z)$ of the SC vortex, is [see
Eq. (\ref{eq3})]:
\begin{equation} \label{hamil}
{\cal H}= {1\over 2{ m}} \left[\vec{p} - q \vec{A}^{(v)}(r,z)
\right]^2 - { 1\over 2} g_{\rm eff} \mu_B\vec{\sigma}\cdot
\vec{B}^{(v)}(r,z)
\end{equation}
where $m$ and $q$ are the effective mass and the charge of the
carrier, and $\vec{A}^{(v)}(r,z)$ is the vector potential,
$\vec{B}^{(v)}(r,z)=\nabla \times \vec{A}^{(v)}(r,z)$.  For
simplicity, we consider a narrow QW, so that the motion is effectively
two-dimensional. As a result, $z$ is just a parameter controlling the
value of the magnetic field. Generalization to a finite width QW has
no qualitative effects.

As discussed in Ref. \onlinecite{berciu03},
for a dipole-like magnetic field such as the one created by the
isolated SC vortex, the eigenfunctions have the general structure:
\begin{equation} \EqLabel{eq8}
\psi_{\rm m}(r, \phi) = \exp{(i{\rm m}\phi)}\left( \begin{array}[c]{c}
\psi_{\uparrow}^{({\rm m})}(r) \\ \psi_{\downarrow}^{({\rm
m})}(r)\exp{[i\phi]} \\ \end{array} \right)
\end{equation}
where ${\rm m}$ is an integer and $\phi=\tan^{-1}(y/x)$ is the polar
angle. The radial equations satisfied by the up and down spin
components can be derived straightforwardly:
$$
\left[\!-\frac{1}{r}\!\frac{d}{dr}\!\left(\!r\frac{d}{dr}\!\right)+\!\frac{{\rm
m}^2}{r^2}- \tilde{g} b_z(r) - e \right]\!\!\psi_\uparrow(r)=
\tilde{g} b_r(r)\psi_\downarrow(r)
$$
$$
\left[\!-\frac{1}{r}\!\frac{d}{dr}\!\left(\!r\frac{d}{dr}\!\right)\!+\!\frac{({\rm m}\!+\!1)^2}{r^2}
\!+\! \tilde{g} b_z(r) \!-\!
e \right]\!\!\psi_\downarrow(r)\!=\!\tilde{g}
b_r(r)\psi_\uparrow(r)
$$
all lengths are in units of $\lambda$. The unit of energy is
$E_0={\hbar^2}/(2m\lambda^2)= 0.05$ meV if we use $m=0.5m_e$ and
$\lambda=40$ nm. $e = E/E_0$ is the eigenenergy. The magnetic fields
have been rescaled, $\vec{B}^{(v)}(r,z) = \phi_0/(4\pi \lambda^2)
\cdot \vec{b}(r)$ (from now on, the distance $z$ between the DMS and
QW is no longer explicitly specified). Finally, $\tilde{g} = g_{\rm
eff}\mu_B\phi_0/(8\pi\lambda^2 E_0)= g_{\rm eff}/8$ if $m=0.5m_e$. The
$\vec{A}^{(v)}(\vec{r})$ terms were left out. This is justified since
they are negligible compared to the Zeeman term, which is enhanced by
$g_{\rm eff}\sim 10^2$ (this was verified numerically). The bound
eigenstates $e <0$ are found numerically, by expanding the up and down
spin components in terms of a complete basis set of functions (cubic
B-splines).

\begin{figure}[t]
\includegraphics[width=\columnwidth]{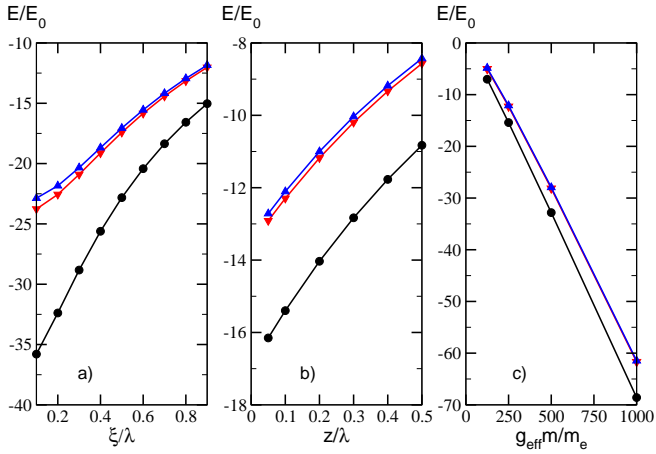}
\caption{ Energy of the
ground state (circles) and of the first two excited states
(triangles), for a charge carrier trapped in the Zeeman potential
created by an isolated vortex, as a function of (a) the coherence
length of the SC; here {\bf $z=0.1 \lambda$}, $m=0.5 m_e$ and $g_{\rm
eff} = 500$; (b) the distance between the DMS layer and the SC. Here
$\xi/\lambda = 35/40$, and the other parameters are as in (a); and (c)
the value of $g_{\rm eff} m/m_e$. Here $\xi/\lambda = 35/40$. In all
cases, $E_0=\hbar^2/(2m\lambda^2)=0.05$ meV, for $\lambda = 40$ nm.}
\label{Fig2b}
\end{figure}

The energies of the ground state (${\rm m}=0$) and the
first two excited trapped states (${\rm m}=\pm 1$) are shown in
Fig. \ref{Fig2b}, as a function of (a) the ratio $\xi/\lambda$, (b)
the distance $z/\lambda$ to the quantum well, and (c) the ratio
$g_{\rm eff}m/m_e$. As expected, the binding energies are largest when
the magnetic fields and therefore the Zeeman potential well are
largest, in the limit $z \rightarrow 0$ and $\xi/\lambda \rightarrow
0$. Since $\xi/\lambda$ controls the spatial extent of the Zeeman
trap, the distance between the ground and first excited states
increase for decreasing $\xi/\lambda$. All binding energies increase
basically linearly with increasing $g_{\rm eff}$.

\begin{figure}[b]
\includegraphics[width=0.9\columnwidth]{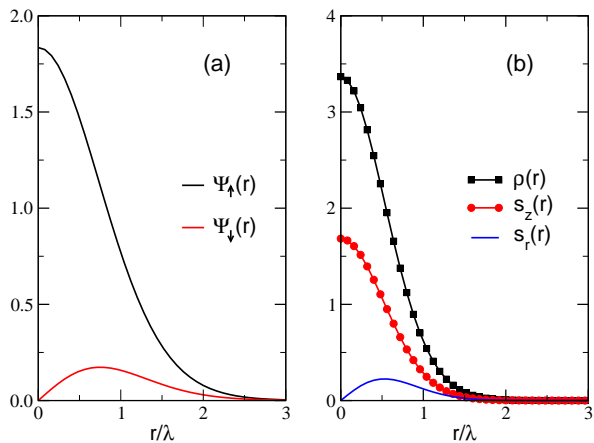}
\caption{ (a)$\psi^{(0)}_{\uparrow}(r)$ and $\psi^{(0)}_{\downarrow}(r)$ of
Eq. (\ref{eq8}), for the ground-state, when $z=0.1 \lambda $(b) The
corresponding density of charge $\rho(r)= |\psi^{(0)}_{\uparrow}(r)|^2
+ |\psi^{(0)}_{\downarrow}(r)|^2$, transversal spin density $s_z(r) =
{1 \over 2} ( |\psi^{(0)}_{\uparrow}(r)|^2 -
|\psi^{(0)}_{\downarrow}(r)|^2 ) $ and radial spin density $s_r(r) =
{\rm Re}(\psi^{(0)*}_{\downarrow}(r) \psi^{(0)}_{\uparrow}(r))$.}
\label{Fig2c}
\end{figure}

The ground-state wavefunction, corresponding to ${\rm m}=0$ in
Eq. (\ref{eq8}), is shown in Fig.~\ref{Fig2c}a. While
$\psi^{(0)}_{\downarrow}(r)\ne 0$ because of the presence of the
radial component, its value is significantly smaller than that of the
spin-up component $\psi^{(0)}_{\uparrow}(r)$ which is favored by the
Zeeman interaction with the large transversal magnetic field. The
expectation value of the trapped charge carrier spin has a
``hedgehog''-like structure, primarily polarized along the $z$-axis,
but also having a small radial component (see Fig.~\ref{Fig2c}b).  The
general structure of these eigenfunctions [Eq. (\ref{eq8})] has been
shown to be responsible for allowing coupling to only one circular
polarization of photons which are normally incident on the DMS layer,
suggesting possible optical manipulation of these trapped, highly
spin-polarized charge carriers \cite{berciu03}. After using quite
drastic analytic simplifications, the results of Ref. \cite{berciu03}
also suggested that depending on the orientation (sign) of the $B_z$
component, only one set of excited state (either ${\rm m>0}$ or ${\rm
m<0}$) is present. Numerically, we find both sets of states present,
with a small lifting of their degeneracy.

\section{Two-dimensional vortex lattice\label{2d}}

\begin{figure}[t]
\includegraphics[width=0.9\columnwidth]{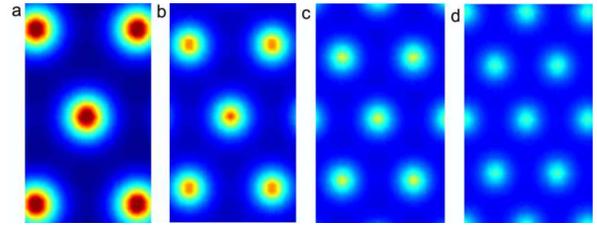}
\caption{Modulation of the transversal component of the magnetic field
$\vec{B}_L(\vec{ r};z)$ of the flux lattice in a given area of a type
II superconductor, for an applied external field $B_0=0.07$ T,
$B_0=0.10$ T, $B_0=0.15$ T and $B_0=0.19$ T respectively in (a), (b),
(c) and (d). As the applied field $B_0$ increases, the distance
between vortices and the modulation of the total field decreases. }
\label{Fig3}
\end{figure}

When placed in a finite external magnetic field $B_{c1} < B_0 <
B_{c2}$, the SC creates a finite density of vortices arranged in an
ordered triangular lattice \cite{abrikosov57}. Since each unit cell
encloses the magnetic flux $B_0 a^2 \sqrt{3}/2= \phi_0/2$ of its
vortex, the lattice constant $a\sim 1/\sqrt{B_0}$ is controlled by the
external field $B_0$, see Fig.~\ref{Fig3}. Consequently, $a$ can be
varied considerably, depending on the ratio $T/T_C$ of the temperature
$T$ to the SC critical temperature $T_C$, which sets the value of
$B_{c1}$, and the ratio $B_0/B_{c1}$.  The corresponding Hamiltonian
for the free carriers in the DMS quantum well, in the presence of a SC
vortex lattice, is given by:
\begin{equation} \EqLabel{1}
{\cal H}={1 \over 2m} \left[\vec{p} +e \vec{A}_L(\vec{r};z) \right]^2-
{1\over 2} g_{\rm eff} \mu_B \vec{\sigma}\cdot \vec{B}_L(\vec{r};z)
\end{equation}
Here, $-e$ is the charge of the charge carriers, assumed to be
electrons. Holes can be treated similarly. We use $\vec{r}=(x,y)$ to
describe the 2D position of the charge carrier inside the narrow (2D)
DMS QW. The magnetic field $ \vec{B}_L(\vec{r})$ created by the
triangular vortex lattice is the sum of the fields created by single
vortices [see Eqs. (\ref{br}), (\ref{bz})]:
\begin{equation}
\vec{B}_{\rm L}(\vec{r};z) = \sum_{\vec{R}}^{}
\vec{B}^{(v)}(\vec{r}-\vec{R}; z) = B_0 \hat{z} + \sum_{\vec{G}\ne 0
}^{} e^{i\vec{G}\cdot \vec{r}} \vec{B}_{\vec{G}}(z) \label{BL}
\end{equation}
where the triangular lattice is defined by $\vec{R} = na (1,0) + m
{a\over 2}(1, \sqrt{3})$, $n, m \in Z$, and $\vec{G}$ are the
reciprocal lattice vectors. The first term is the average field per
unit cell, which equals the applied external field $B_0\hat{z}$. The
second term is the periodic field induced by the screening
supercurrents.  This term has zero flux through any unit cell and
decreases rapidly as the distance $z$ between the SC and the DMS
layers increases. As in the previous section, $z$ is here just a
parameter, and we will not write it explicitly from now on.

Likewise, we separate the vector potential in two parts, corresponding to each
contribution of the magnetic field: $$ \vec{A}_{\rm
L}(\vec{r})=\vec{A}_0(\vec{r}) + \vec{a}(\vec{r}).$$ In the Landau
gauge, $\vec{A}_0(\vec{r})=(0, B_0 x, 0)$ and $ \vec{a}(\vec{r})=
\sum_{\vec{G}\ne 0 }^{}e^{i \vec{G}\cdot \vec{ r}} \vec{a}_{\vec{G}}$,
with $ a_{\vec{G}} = [i \vec{G}\times
\vec{B}_{\bar{G}}(z)]/|\vec{G}|^2$.

Before constructing the solutions for the full Hamiltonian of
Eq. (\ref{1}), we first briefly review the
solutions in the presence of only a homogeneous field $B_0$, in order
to fix the notation. In this case the Zeeman term lifts the spin
degeneracy, but it is the orbital coupling that essentially determines
the energy spectrum of the carriers, which consists of spin-polarized
Landau Levels (LL):
\begin{equation}
E_{N,\sigma} =\hbar\omega_c\left(N+{1\over 2}\right) - {1\over 2}
g_{\rm eff} \mu_B B_0\sigma.
\end{equation}

In the Landau gauge the corresponding
eigenstates are:
 \begin{equation}
\phi_{N,k_y,\sigma}(\vec{r})={e^{ik_yy}e^{-{1\over2}
\left( {x\over l}  + lk_y\right)^2}\over \sqrt{b}\sqrt{l\sqrt\pi 2^N
N!}}  H_N\left({x\over l} + lk_y\right)
\chi_\sigma, \label{phill}
\end{equation}
 where $l = \sqrt{\hbar/(eB_0)}$ is the magnetic length and $\omega_c
= eB_0/m$ is the cyclotron frequency.  $N\ge 0$ is the index of the
LL, $H_N(\zeta)$ are Hermite polynomials and $k_y$ is a
momentum. $\chi_\sigma$ are spin eigenstates, $\sigma_z \chi_\sigma =
\sigma \chi_\sigma $, $\sigma = \pm 1$. Each LL is highly degenerate
and can accommodate up to one electron per $2\pi l^2$ sample area. The
filling factor $\nu = n 2\pi l^2$, where $n$ is the 2D electron
density in the DMS QW, counts how many LL are fully or partially
filled.

This spectrum of highly degenerate LLs is very different from
that arising if only the {\em periodic} part of the magnetic field and
vector potentials are present in the Hamiltonian. In this case, the
system is invariant to the discrete lattice translations. As a result,
one finds the spectrum to consist of electronic bands defined in a
Brillouin zone determined by the periodic Zeeman potential, the
eigenfunctions being regular Bloch states. One can roughly think of
these states as arising from nearest-neighbor hopping between the
states trapped under each individual vortex, discussed in the previous
section.

In order to construct solutions that include both the
periodic and the homogeneous part of the magnetic field, we need to
first consider the symmetries of Hamiltonian (\ref{1}). Because of the
orbital coupling to the non-periodic part of the vector potential
$\vec{A}_0$, the ordinary lattice translation operators $T(\vec{R}) =
e^{{i \over \hbar} \vec{R}\cdot\vec{p} } $ do not commute with the
Hamiltonian. Instead, one needs to define so-called magnetic
translation operators~\cite{zak64}:
$$ T_M(\vec{R}) = e^{-{ie \over
\hbar} B_0 R_x y} e^{{i \over \hbar} \vec{R}\cdot\vec{p} }. $$
It is straightforward to verify that these operators commute with the
Hamiltonian, $[{\cal H}, T_M(\vec{R})]=0$, and that
$$ T_M(\vec{R})T_M(\vec{R'})= e^{-{ie \over \hbar} B_0
R'_xR_y}T_M(\vec{R}+\vec{R'}). $$

\begin{figure}[t]
\includegraphics[width=0.9\columnwidth]{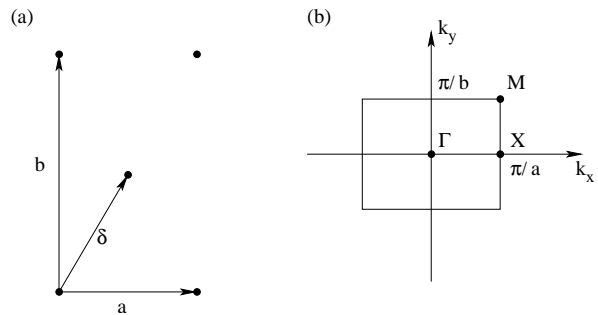}
\caption{\label{Fig5} (a) Rectangular magnetic unit cell containing 2
vortices, and its lattice vectors. (b) The associated magnetic
Brillouin zone and the location of the special, high-symmetry points
$\Gamma$, X and M.}
\end{figure}

It follows that these operators form an Abelian group provided that we
define a {\em magnetic} unit cell so that the magnetic flux through it
is an integer multiple of $\phi_0$. In our case, as already discussed,
the magnetic flux through the unit cell of the vortex lattice is
precisely $\phi_0/2$. We therefore define the magnetic unit cell to be
twice the size of the original one.  We use a rectangular unit cell,
as shown in Fig.~\ref{Fig5}a. The new lattice vectors are
$\vec{a}=(a,0)$ and $\vec{b}=(0,b)$, with $b=a\sqrt{3}$, and
$\vec{R}_{nm} = n \vec{a}+m\vec{b}$. The basis consists of two
vortices, one placed at the origin and one placed at
$\vec{\delta}=(\vec{a}+\vec{b})/2$. The associated magnetic Brillouin
zone is $k_x \in (-\pi/a,\pi/a]; k_y \in (-\pi/b,\pi/b]$ and the
reciprocal magnetic lattice vectors are $\vec{G}_{n,m} = n (2\pi/a)
\hat{x} + m (2\pi/b) \hat{y}$, $n,m \in Z$ (see Fig. \ref{Fig5}b).

With this choice, the wave-functions must also be eigenstates of the
magnetic translation operators:
\begin{equation} \label{tm}
T_M(\vec{R}) \psi_{\vec{k}}(\vec{r}) = e^{i\vec{k}\cdot \vec{R}}
\psi_{\vec{k}}(\vec{r}).
\end{equation}
We thus need to expand the eigenstates in a complete basis set of
wavefunctions which satisfy Eq. (\ref{tm}). Such a basis can be
constructed from the LL eigenstates, since the large degeneracy of
each LL allows one to construct linear combinations satisfying
Eq. (\ref{tm})~\cite{thouless82}:
\begin{equation} \label{s45}
\Psi_{N,\vec{k},\sigma}(\vec{r}) = { 1 \over \sqrt{N_T}
}\sum_{n=-\infty}^{\infty} e^{ik_x na} \phi_{N,k_y+{2\pi \over
b}n,\sigma}(\vec{r})
\end{equation}
where $N_T$ is the number of magnetic unit cells and $\vec{ k}$ is a
wavevector in the magnetic Brillouin zone.

As a result, we search for eigenstates of Hamiltonian (\ref{1}) of the
general form:
\begin{equation}  \label{s46}
 \psi_{\vec{k}}(\vec{r}) = \sum_{N,\sigma} d_{N\sigma}(\vec{k})
\Psi_{N,\vec{k},\sigma}(\vec{r})
\end{equation}
 Here, $d_{N\sigma}(\vec{k})$ are complex coefficients characterizing
 the contribution of states from various LLs to the true
 eigenstates. Spin-mixing is necessary because $[{\cal
 H},\hat{\sigma_z}] \ne 0$.

 Note that since the periodic potential
 may be large (due to the large Land\'e factor) we cannot make the
 customary assumption that it is much smaller than the cyclotron
 frequency, and thus assume that there is no LL mixing
 \cite{thouless82,pfannkuche92,demikhovskii03b,usov88}. Instead, we
 mix a large number of LLs, so that we can find the exact solutions
 even in the case when the cyclotron frequency is much smaller than
 the amplitude of the periodic potential.

 The problem is now reduced
 to finding the coefficients $d_{N\sigma}(\vec{k})$, constrained by
 the normalization condition $\sum_{N,\sigma}^{}
 |d_{N\sigma}(\vec{k})|^2 = 1$. The Schr\"odinger equation reduces to
 a system of linear equations:
\begin{equation}  \EqLabel{s47}
\left[ E_{\vec{k}}-\!E_{N,\sigma}\right] \! d_{N\sigma}(\vec{k})
\!=\! -{g_{\rm eff} \mu_B\over 2} \!\! \sum_{N'\sigma'}^{}
d_{N'\sigma'}(\vec{k}) {\vec \sigma}_{\sigma\sigma'}\cdot
\vec{b}_{N,N'}
\end{equation}
 where $$ \vec{b}_{N,N'}=\int_{}^{} d
\vec{r} \Psi^*_{N,\vec{k}}(\vec{r})\left[\vec{B}_{\rm L}(\vec{r};z)
-B_0 \hat{z}\right]\Psi_{N',\vec{k}}(\vec{ r}) $$
 $$=\sum_{\vec{G}\ne
0}^{}\vec{B}_{\vec{ G}}(z)\exp[{-i(k_xG_y-k_yG_x)l^2}]I_{N,N'}^{\vec{G}}$$
Here (see Ref. ~\onlinecite{pfannkuche92}):
$$ I_{N,N'}^{\vec{G}}
=\sqrt{{m! \over M!}} \!\left(i \sqrt{\tilde{G}}\right)^{\!M-m}
\!\!\!\!\!e^{-{\tilde{G} \over 2}} L_m^{\!M-m}(\tilde{G}) \left[{ G_x
- iG_y \over G}
\right]^{\!N-N'} $$
 where $m = \min(N,N')$, $M = \max(N,N')$,
$\tilde{G}=l^2G^2/2$. $L_n^m(x)$ are associated Laguerre
polynomials. The Fourier components $\vec{B}_{\vec{G}}(z)$ of the
magnetic field can be calculated straightforwardly, see
Eqs. (\ref{br}), (\ref{bz}) and (\ref{BL}). For the reciprocal vectors
of the magnetic unit cell, we find that $\vec{B}_{\vec{G}_{nm}}(z) =0$
if $n+m$ is an even number, otherwise:
$$ \vec{B}_{\vec{G}}(z) =
(-iG_x, -iG_y, G) \cdot \frac{B_0 e^{-Gz-\frac{1}{2}G^2\xi^2}}{G\tau
(G\lambda+\tau)} $$
where $\tau = \sqrt{G^2 \lambda^2 + 1}$.

Eq. (\ref{s47}) neglects the periodic terms proportional to
$\vec{a}(\vec{r};z)$. These are small compared to the periodic terms
related to $\vec{B}_{\rm L}(\vec{r};z)$, since the latter are
multiplied by $g_{\rm eff}\gg 1$ (we verified this explicitly). We
numerically solve Eq. (\ref{s47}), typically mixing LL up to $N=40$
and truncating the sum over reciprocal lattice vectors in
$\vec{b}_{N,N'}$, to the shortest 1600. These cutoffs are such that
the lowest bands eigenstates $\psi^{(\alpha)}_{\vec{k}}(\vec{r})$ and
dispersion $E^{(\alpha)}_{\vec{k}}$ ($\alpha$ is the band index) are
converged and do not change if more LL and/or reciprocal lattice
vectors are included in the calculation.

\begin{figure}[t]
\includegraphics[width=0.95\columnwidth,clip]{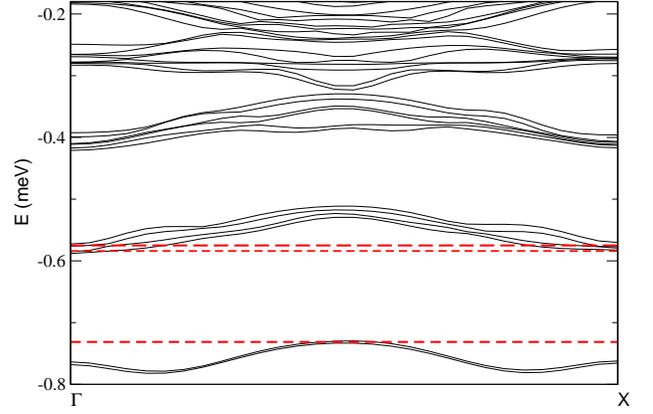}
\caption{\label{Fig5a} The energy spectrum in the lattice case with
$a=8.5\lambda$ (solid lines) compared to lowest energy eigenstates of
the isolated vortex (dashed lines).  SC parameters correspond to Nb
and $z=0.1\lambda$.}
\end{figure}

A first test for our
numerical results is to compare results obtained in the lattice limit
$B_0\rightarrow 0$, $a \rightarrow \infty$, with the energy spectrum
of the isolated vortices, obtained in the previous section. One such
comparison is shown in Fig.~\ref{Fig5a}, where the band dispersion for
a lattice with $a=8.5\lambda$ is shown to have the location of the
lowest bands in good agreement with the eigenenergies of the isolated
vortex. The agreement improves for larger $a$ values.

Another check on our results is to investigate the dispersion of the
lowest-energy bands, in this limit. As discussed before, in the
absence of the orbital coupling to the $\vec{A}_0$ component of the
vector potential, one expects a simple tight-binding dispersion for
the lowest bands, with an effective hopping $t$ characterizing the
overlap between eigenfunctions trapped under neighboring vortexes. In
the presence of the orbital coupling, the hopping matrices pick up an
additional phase phase factor proportional to the enclosed flux (the
Peierls prescription). This is responsible for lifting the degeneracy
of each tight-binding band. The resulting sub-band structure, and in
particular the number of sub-gaps opened in each tight-binding band,
are known to depend only on the ratio $\phi/\phi_0$, where $\phi$ is
the flux of the applied magnetic field through the unit cell of the
periodic potential. In fact, for $\phi/\phi_0=p/q$, where $p$ and $q$
are mutually prime integers, each tight-binding band splits into
precisely $q$ subbands~\cite{hofstadter76}. This problem has been well
studied as one of the asymptotic limits of the Hofstadter butterfly,
corresponding to a large periodic modulation and a small applied
magnetic field.
\begin{figure}[t]
\includegraphics[width=0.95\columnwidth,clip]{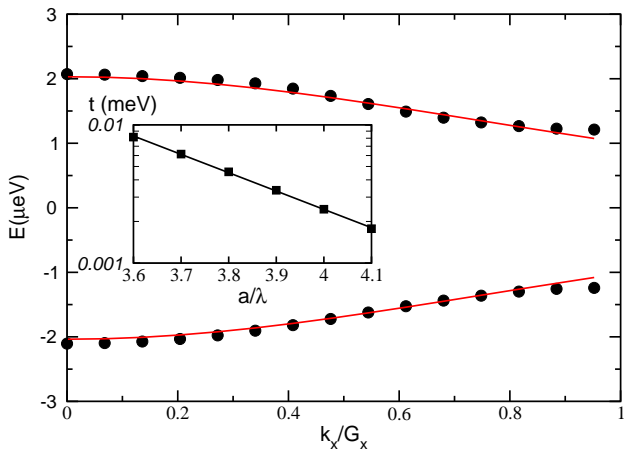}
\caption{\label{Fig5b}Fit of the energy spectrum of the two lowest
energy sub-bands (circles) with the expressions of Eq. (\ref{thof})
appropriate in the asymptotic limit $a \gg \lambda$ (lines). This
energy spectrum corresponds $B_x=B_y=0$ (see text) and $a=4\lambda$,
and it is shown as a function of $k_x$, with $k_y=0$.  Its overall
energy values have been shifted for convenience. The inset shows the
effective hopping $t$ extracted from this fit, which decreases
exponentially with increasing distance between vortices.}
\end{figure}

In our case, $\phi/\phi_0=1/2$ and we expect each tight-binding band
to split into two sub-bands. This is indeed verified in all the
asymptotic limits $a \gg \lambda$ where the tight-binding
approximation is appropriate (in some of the plots that we show, such
as Fig. \ref{Fig5a}, this splitting is too small for the lowest band
and is not visible on this scale). In fact, we can even fit the
dispersion of the bands, in this asymptotic limit. For a triangular
lattice with nearest neighbor hopping $t$ (real number), it is
straightforward to show that the dispersion when a magnetic field with
$\phi/\phi_0=1/2$ is added is (see Ref. \onlinecite{hatsugai90}):

\begin{equation}
E(k_x,k_y)=\pm2t\sqrt{1+\cos^2(k_x a)-\cos(k_x
a)\cos(k_y b)}.
\label{thof}
\end{equation}

In our case, the hopping
$t$ between states trapped under neighboring vortices is not a real
number, even if we set $\vec{A}_0(\vec{r})=0$ . The reason is that the
wave-functions have a non-trivial spinor structure (see
Eq. (\ref{eq8})) which leads to a complex value of $t$ ($t$ is just a
matrix element related to the overlap of neighboring
wavefunctions). To avoid complications coming from dealing with the
phase of $t$ and the changes induced by it on the simple dispersion of
Eq. (\ref{thof}), we test the fit for a ``toy model'' in which we set
the in-plane magnetic field to zero: $B_x(x,y)=B_y(x.y)=0$. In this
case, the spin is a good quantum number, the ground-state
wavefunctions are simple $s$-type waves, the corresponding hopping $t$
is real and Eq. (\ref{thof}) holds. We show a fit for such a case in
Fig.~\ref{Fig5b}, along the $k_y=0$ line in the Brillouin
zone. This fit allows us to extract a value for the effective hopping
$t$. As expected (see inset of Fig.~\ref{Fig5b}), $t$ decreases
exponentially with increasing distance $a$ between neighboring
vortices.

\begin{figure}[h]
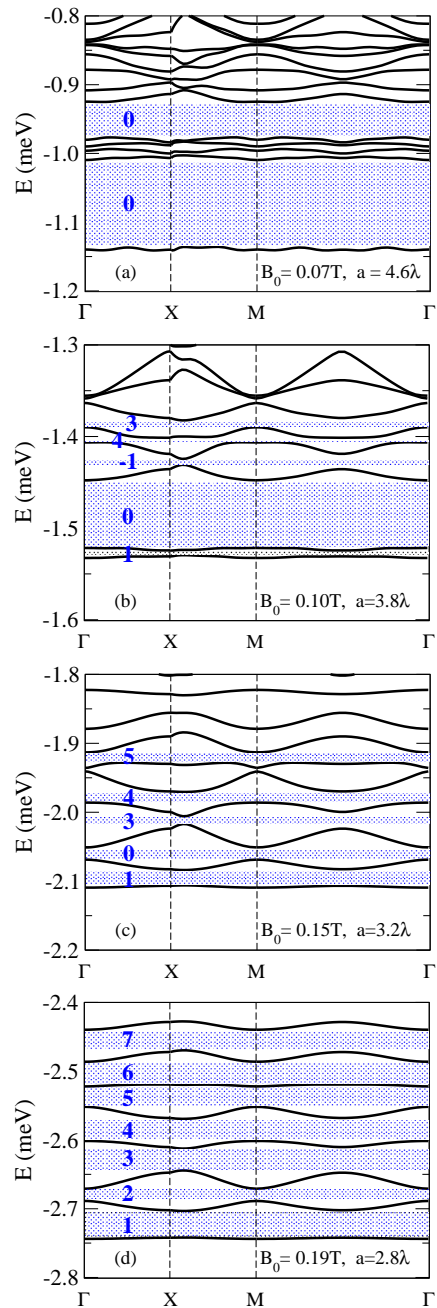

\includegraphics[width=0.65\columnwidth]{FIG10a.eps}
\includegraphics[width=0.65\columnwidth]{FIG10b.eps}
\includegraphics[width=0.65\columnwidth]{FIG10c.eps}
\includegraphics[width=0.65\columnwidth]{FIG10d.eps}
\caption{Band structure for the carriers in the DMS QW for $B_0$ of
(a) $0.07T$;(b) $0.10T$; (a) $0.15T$;(a) $0.19T$. The gaps are marked
by shaded regions, and each is labelled with its Chern number (see
text). }\label{Fig4}
\end{figure}

The presence of the in-plane components changes the structure of the
wavefunctions and leads to complex values of $t$, modifying the
dispersion from the simple Eq. (\ref{thof}) form. Indeed, as one can
see from Fig. \ref{Fig5a}, the dispersion is now quite different in
shape than the one shown for the toy model in Fig. \ref{Fig5b}. In
fact, although there are two distinct subbands $E(k_x,k_y)$ for any
point in the Brillouin zone, their gaps do not overlap and so there is
no true subgap appearing in the band-structure.

These results, corresponding to the asymptotic limit of a large
modulation and small applied field, show that our numerical
scheme based on expansion in
terms of multiple LLs is working well even in this most unfavorable
limit. The other asymptotic limit where our results can be easily
verified against known predictions is the limit of a large applied
field and small periodic modulation. In this case, the small
modulation is expected to lift the degeneracy of each LL, but not to
lead to mixing amongst the LL. This problem has also been studied
extensively~\cite{claro79,thouless82,pfannkuche92} and the resulting
spectrum is also known to depend only on the ratio
$\phi/\phi_0=p/q$. Unlike in the tight-binding limit, here each LL
splits into $p$ subbands.

In our case, $p=1$ and therefore we expect
no supplementary structure in the LLs. This is indeed verified, as
shown, for example, in Fig.~\ref{Fig4} where we plot the evolution of
the electronic band structure as $a$ is varied. For $a \gg \lambda$
(panel (a)) we see the emergence of the tight-binding structure
discussed above. As $a$ decreases with increasing $B_0$ (panel (d)),
we indeed see the emergence of nearly equidistant Landau levels, which
still exhibit some dispersion due to the weak periodic
potential. Because of the large, quasi-uniform Zeeman interaction in
this limit, all these states are mostly spin-up and the splitting
between consecutive bands corresponds to the cyclotron energy.  The
intermediary cases (panels (b) and (c)) correspond to situations where
neither asymptotic limit is appropriate. In such cases one needs to
perform numerical simulations to find the resulting band-structure.
The most direct signature of this strongly field-dependent
band-structure is obtained in magnetotransport measurements, which we
proceed to discuss now.

\section{Integer Quantum Hall Effect\label{IQHE}}

Following the discovery of the integer quantum Hall effect in a two
dimensional electron gas (2DEG) in a strong magnetic field, Laughlin
demonstrated that the Hall conductance of a non interacting 2DEG is a
multiple of $e^2/h$ if the Fermi energy lies in a mobility gap between
two LL~\cite{laughlin81}. Later, Thouless {\it et al.} argued that
the quantization of the Hall conductivity $\sigma_{xy}$ in
periodically modulated systems has a topological nature and therefore
it occurs whenever the Fermi energy lies in an energy gap, even if the
gap lies within a Landau level~\cite{thouless82}. Using the Thouless
formula, which we discuss below, the Hall conductances corresponding
to various types of lattices, in either of the two asymptotic limits
of strong or weak periodic potential, have then been
investigated~\cite{springsguth97}.

In our case, it is necessary to
use a more general method to calculate the Hall conductance. We
briefly review it here. This approach is inspired by the work of
Kohmoto\cite{kohmoto85} and Usov~\cite{usov88}, who showed that it is
possible to use the topology of the band-structure to calculate, in a
relatively simple way, the contribution to the Hall conductance of
each filled subband.

When the Fermi level is inside a gap, the total Hall conductivity
equals the sum of the contributions from all the fully occupied
bands: $\sigma_{xy} = \sum_{\alpha} \sigma_{xy}^{(\alpha)}
\Theta(E_F - E^{(\alpha)}_{\vec{k}})$. The contribution of each
occupied band is~\cite{thouless82}:
\begin{equation}
 \sigma^{(\alpha)}_{xy} = i { e^2 \over 2\pi h}
\int_{}^{}d\vec{k}\int_{}^{}d\vec{r} \left[\frac{\partial
u_{\vec{k}}^{(\alpha)*}}{\partial k_x} \frac{\partial
u_{\vec{k}}^{(\alpha)}}{\partial k_y} -\frac{\partial
u_{\vec{k}}^{(\alpha)*}}{\partial k_y} \frac{\partial
u_{\vec{k}}^{(\alpha)}}{\partial k_x} \right]
\end{equation}
where the integrals are carried over the magnetic Brillouin zone and
the magnetic unit cell, respectively. Here,
$u_{\vec{k}}^{(\alpha)}(\vec{r})=\psi^{(\alpha)}_{\vec{k}}
(\vec{r})e^{-i\vec{k}\cdot\vec{r}}$ is the Bloch part of the band
eigenstate. Using Eqs. (\ref{s45}) and (\ref{s46}) we can perform
the real space integrals explicitly to obtain:
\begin{equation}
\sigma^{(\alpha)}_{xy} = {e^2\over h} \left[1+ { 1 \over 2\pi}
\mbox{Im}\int_{}^{}d\vec{k}\sum_{N\sigma} \frac{\partial
d_{N\sigma}^{(\alpha)*}(\vec{k}) }{\partial k_y} \frac{\partial
d_{N\sigma}^{(\alpha)}(\vec{k}) }{\partial k_x} \right].
\end{equation}
The first term is the unit conductance contribution of each LL
(band), expected in the absence of the periodic modulation -- the
usual IQHE. The second term can be rewritten as
\begin{equation}
\Delta\sigma^{(\alpha)}_{xy}={e^2\over h 2\pi i}
\int_{}^{}d\vec{k}\hat{e_z}\cdot \left[\nabla_{\vec{k}} \times
\vec{A}^{(\alpha)}(\vec{k})\right]
\end{equation}
where
\begin{equation}
\vec{A}^{(\alpha)}(\vec{k}) = \sum_{N\sigma}^{}
d_{N\sigma}^{(\alpha)}(\vec{k}) \nabla_{\vec{k}}
d_{N\sigma}^{(\alpha)*}(\vec{k}).
\end{equation}
This term can be calculated using Kohmoto's
arguments~\cite{kohmoto85}. The magnetic Brillouin zone is a $T^2$
torus and has no boundaries. As a result, if the gauge field
$\vec{A}^{(\alpha)}(\vec{k})$ is uniquely defined everywhere,
Stoke's theorem shows that this integral vanishes. The gauge field
is related to the global phase of the band eigenstates: if
$\psi^{(\alpha)}_{\vec{k}}(\vec{r})\rightarrow e^{i
f(\vec{k})}\psi^{(\alpha)}_{\vec{k}}(\vec{r})$, then
$\vec{A}^{(\alpha)}(\vec{k}) \rightarrow \vec{A}^{(\alpha)}(\vec{k})
- i \nabla_{\vec{k}}f(\vec{k})$. It follows that the gauge field can
be uniquely defined if the phase of any one of the components
$d_{N\sigma}^{(\alpha)}(\vec{k})$ (and thus, the global phase of the
band eigenfunction) can be uniquely defined in the entire magnetic
Brillouin zone. This is possible only if the chosen component
$d_{N\sigma}^{(\alpha)}(\vec{k})$ has no zeros inside the magnetic
Brillouin zone, in which case we can fix the global phase by
requesting that this component be real everywhere. In this case, as
discussed, $ \sigma^{(\alpha)}_{xy} =e^2/h$.  If there is at least
one point $\vec{k}_0$ where $d_{N\sigma}^{(\alpha)}(\vec{k}_0)=0$,
in its vicinity the global phase must be defined from the condition
that some other component $d_{N'\sigma'}^{(\alpha)}(\vec{k})$, which
is finite in this region, is real. The definitions of the global
phase inside and outside this vicinity of $\vec{k}_0$ are related
through a gauge transformation; moreover, the torus now has a
boundary separating the two areas. Applying Stokes' theorem, it
follows immediately~\cite{kohmoto85} that
$\Delta\sigma^{(\alpha)}_{xy}={e^2 \over h} S_0$. The integer $S_0$
is the winding number in the phase of
$d_{N'\sigma'}^{(\alpha)}(\vec{k})$ (when
$d_{N\sigma}^{(\alpha)}(\vec{k})$ is real) on any contour
surrounding $\vec{k}_0$. If $d_{N\sigma}^{(\alpha)}(\vec{k})$ has
several zeros inside the Brillouin zone, then one has to sum the
winding numbers associated with each zero:
\begin{equation}
\sigma^{(\alpha)}_{xy}=\frac{e^2}{h}\left[ 1+ \sum_m S_m\right]
\end{equation}
Thus, once the coefficients $d_{N\sigma}^{(\alpha)}(\vec{k})$ are
known, one can immediately find the contribution of the band
$\alpha$ to $\sigma_{xy}$ by studying their zeros and their
vorticities.
\begin{figure}[t]
\includegraphics[width=0.9\columnwidth]{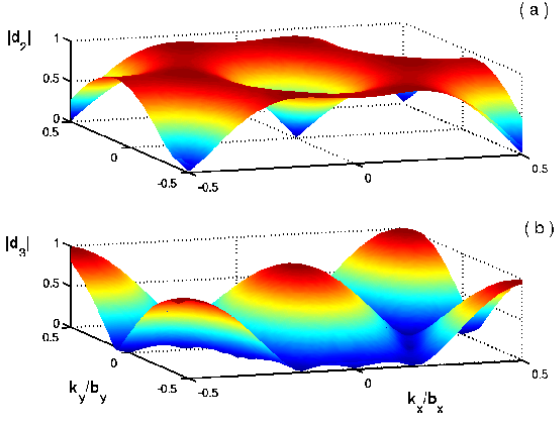}
\caption{\label{Fig7} Values in the magnetic Brillouin zone of the
$d^{(\alpha)}_{N,\sigma}(\vec{k})$ coefficients corresponding to the
second band $\alpha=2$, in the band-structure obtained for $B_0=0.095
T $, $z=8$ nm and $g=500$.  a) $d^{(2)}_{2,\uparrow}(\vec{k})$ is
chosen to be real everuwhere. It has zeros in the corners and center
of the magnetic Brillouin zone. b) $|d^{(2)}_{3,\uparrow}(\vec{k})|$
has maxima where $d^{(2)}_{2,\uparrow}(\vec{k})$ has zeroes.}
\end{figure}

As an illustration of this method, we calculate the contribution of
the second band $\alpha=2$, for $B_0=0.095 T $, $z=8$ nm and
$g=500$. We first choose the overall phasefactor for the corresponding
coefficients $d^{(2)}_{N,\sigma}(\vec{k})$ defining this band so that
one of them (specifically, here we chose $N=2$) is real throughout the
Brillouin zone, see Fig. \ref{Fig7}. We see that this component has
zeros in some high-symmetry points, signalling a potentially
non-trivial contribution to $\sigma_{xy}$. In order to find the
corresponding vorticities, we investigate any other component that has
no zeros at the positions of the singular points of
$d^{(2)}_{2,\uparrow}$. In general, this other
component will have complex values. In the lower panel of
Fig. \ref{Fig7} we plot the absolute value of $d^{(2)}_{3,\uparrow}$,
which is indeed finite at all zeros of $d^{(2)}_{2,\uparrow}$.

\begin{figure}[t]
\includegraphics[width=0.75\columnwidth]{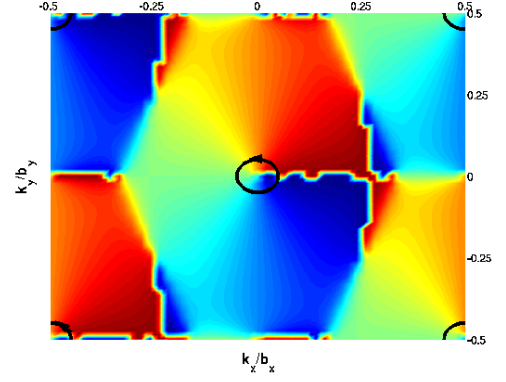}
\caption{\label{Fig8} Phase map of $d^{(2)}_{3,\uparrow}$ inside the
MBZ. Blue represents $-\pi$ and red represents $\pi$. The phase winds
by $-2\pi$ around both the center and the corner points.}
\end{figure}

The vorticities can be found by investigating the variation in the phase
of $d^{(2)}_{3,\uparrow}$ around the zeros of
$d^{(2)}_{2,\uparrow}$. The phase-map of $d^{(2)}_{3,\uparrow}$ is
shown in Fig.~(\ref{Fig8}), where red represents a phase of $\pi$ and
blue a phase of $-\pi$. Any boundary red-blue indicated a non-zero
winding number. These winding numbers $S_m$ (in units of 2$\pi$) are
the phase incursions of $d^{(2)}_{3,\uparrow}$ upon going
anticlockwise around the singular points. In this case, it is clear
that the phase incursions are of $-2\pi$. As there are two singular
points in the MBZ (the corners count as 1/4 each), the conductance of
the subband 2 is
$\sigma^{(2)}_{xy}=(\frac{e^2}{h})(1-2)=-(\frac{e^2}{h})$. The same
overall value is obtained by looking at the vorticities of any other
wavefunction that is finite at the zeros of $d^{(2)}_{2,\uparrow}$.

\begin{figure}[b]
\includegraphics[width=0.75\columnwidth]{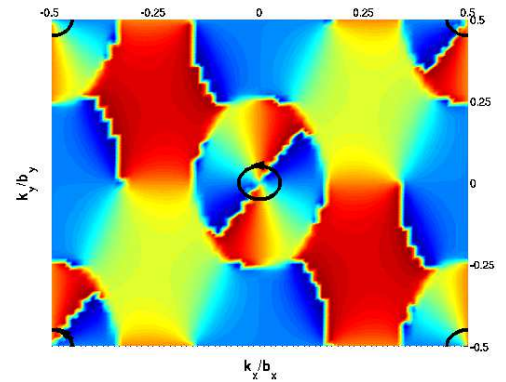}
\caption{\label{Fig9} Same as in Fig. \ref{Fig8} but for component
$d^{(6)}_{8,\uparrow}(\vec{k})$.}
\end{figure}

Larger winding numbers are also possible. As a second example, in
Fig.~\ref{Fig9} we show the phase map for the component
$d^{(6)}_{8,\uparrow}$ for the sixth band. In this case, we requested
that the component $d^{(6)}_{6\uparrow}$ be real, and we found its
zeros to be again in the center and corners of the Brillouin
zone. Now, the phase incursion about the zeros are $-2\times2\pi$ and
therefore the conductance of the 6$^{th}$ subband is, in this case,
$\sigma^{(6)}_{xy}=(\frac{e^2}{h})(1-4)=-3(\frac{e^2}{h})$. These
integers are called Chern numbers.

As discussed by Avron {\it et al.}~\cite{avron83}, based on the
topological nature of the Hall conductance it can be shown that the
Chern number of a subband does not change unless the subband merges
with a neighboring subband. In this case, the Chern number of the
newly formed band equals the sum of the Chern numbers of the two
subbands. Similarly, if a band splits into one or more subbands, its
Chern number is re-distributed over the two or more subbands. In the
system discussed here, a variation of the applied magnetic field $B_0$
changes the band structure, opening and closing energy gaps. By
calculating the energy spectrum of the system for a given range of the
magnetic field, we map the opening and closing of the gaps between the
few lowest mini-bands as a function of $B_0$ (see Fig. \ref{Fig4}). We
than calculate the Hall conductance of each of the lowest seven
mini-band for each configuration of gaps using this method.

It is important to point out that we recover the expected values of the Hall
conductance of the minibands in both limits of strong and weak
modulation for a triangular lattice and $p/q=1/2$. In Fig. \ref{Fig4},
we see in panel (b) that the first two jumps in the conductivity are
$1 e^2/h$ and then $-1 e^2/h$, as expected for the $p=2$ subbands of
tight-binding limit of the triangular Hofstadter
butterfly~\cite{hatsugai90}. In the limit of weak modulation, we obtain
the usual IQHE, as expected, since $p=1$.

\begin{figure}[t]
\includegraphics[width=0.85\columnwidth]{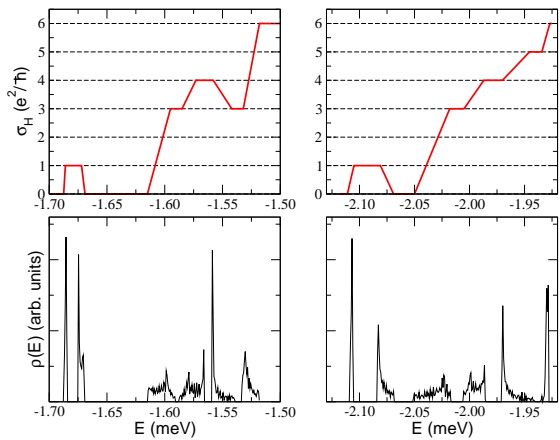}
\caption{\label{Fig10}Density of states and hall
conductivity as a function of the Fermi Energy for two different
applied magnetic fields B$_0$. }
\end{figure}

Such a calculation allows us to predict the sequence of plateaus in
the IQHE if one keeps $B_0$ fixed and varies the concentration of
electrons in the DMS (for instance, by varying the value of a
back-gate voltage). Such predictions are shown in Fig. \ref{Fig10},
where we plot the density of states (lower panels) and the
corresponding Hall conductivity (upper panels) as a function of the
Fermi energy for band-structure corresponding to two different values
of $B_0$. Of course, one needs disorder in order to observe the
IQHE. We did not include disorder in this calculation, but we know
from Laughlin's arguments that the value of the Chern numbers remains
the same, in its presence. The plots in the upper pannels of
Fig. \ref{Fig10} are thus only sketches of the expected Hall
conductances, in these cases. The main observation is that the Hall
conductivity does not increase monotonically as a function of $E_F$,
as is the case in the usual IQHE. This, of course, is due to the
presence of the periodic potential induced by the vortex lattice, and
clearly signals the formation of the Hofstadter butterfly.

\begin{figure}[tb]
\includegraphics[width=0.95\columnwidth,clip]{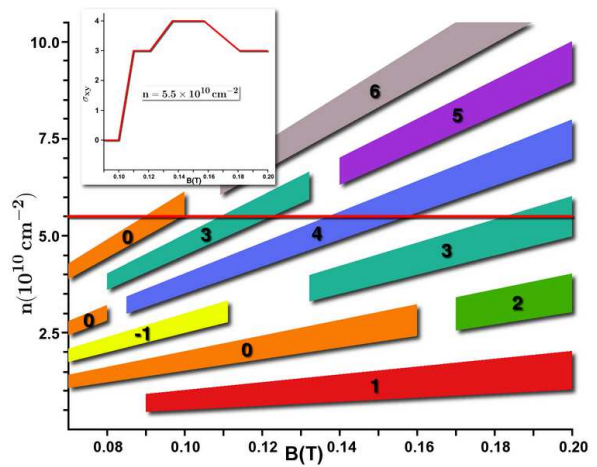}
\caption{conductivity $\sigma_{xy}$ in units of $e^2/h$, as a function
of the magnetic field $B$ and the charge carrier density $n$. The
colored areas mark the gaps between the bands. For some ranges of $B$,
neighboring bands touch and some of the gaps close. The integers give the quantized values of the
$\sigma_{xy}$ plateaus. For example,
$\sigma_{xy}$ as a function of $B$, at a constant density n$=5.5
\times 10^{10}$cm$^{-2}$ (red line) is shown in the inset. It is
quantized every time the Fermi level is inside a gap. The parameters
are the same as for Fig. \ref{Fig3}.}
\label{fignew} \end{figure}

We can summarize more efficiently the information shown in plots like
Fig. \ref{Fig10} in the following way. The two important parameters
are the values of the electron concentration when the Fermi level is
in a gap, and the value of $\sigma_{xy}$ expected for that gap. Such a
plot is shown in Fig. \ref{fignew}. The various colored regions are
centered on the values of the electron concentrations where the Fermi
energy should be in a gap. These regions are assigned an arbitrary
width (calculations involving realistic disorder are needed to find
the widths of these plateaus) and are labelled with the integer $i$
defining the quantized value of $\sigma_{xy}= i (e^2/h)$. As the
external magnetic field is varied, some subbands merge and then
separate, changing their individual Chern numbers in the process. This
plot allows us to predict the sequence of Hall plateaus for a constant
value of the electron density $n$ and varying external field, as shown
in the inset for $n = 5.5 \times 10^{10}$ cm$^{-2}$. The non-monotonic
sequences of plateaus as $B_0$ decreases clearly signals the
appearance of tight-binding bands, which in turn show that the Zeeman
potential is strong enough to localize spin-polarized charge carries
under individual vortices.

\section{Concluding remarks}
In this paper we presented detailed numerical and analytical
calculations aimed at investigating the novel spin and charge
properties of a magnetic semiconductor quantum well in close
proximity of a superconducting flux line lattice. First we have
shown how the single superconducting vortex localizes spin polarized
states, with binding energies within the accessible range of several
local spectroscopic probes, as well as transport measurements. We
then turned our attention to the case of a periodic flux line
lattice, and presented the results of a numerical framework able to
interpolate between the inhomogeneous (low) field regime of dilute
vortex lattice and the homogeneous (high) field regime characterized
by Landau level quantization. Our numerical scheme not only
reproduces the energy spectrum of the isolated vortex limit, but the
spin-polarized electronic band structure we obtain within this
framework also matches with the analytical tight binding
calculations applied for the dilute vortex lattice limit as well.
Between the two extreme field limits we investigated the
momentum-space topology of the Bloch wave functions associated with
the spin polarized bands. By using the connection between the wave
function topology and quantum Hall conductance, we showed how the
consequences of the 1/2 Hofstadter butterfly spectrum can lead to
experimentally observable effects, such as a non-monotonic
'staircase' of Hall plateaus as they appear under  a varying
magnetic field or carrier concentration. We paid special attention
to provide realistic systems and materials parameters for each
experimental configuration we suggested. All indications from our
theory seem to suggest that the magnetic
semiconductor-superconductor hybrids can be fabricated with
presently available molecular-beam epitaxy will provide us with a
rich variety of transport and spectroscopic phenomena.

\section{Acknowledgements}
We would like to thank A. Abrikosov, J.K. Furdyna, T. Wojtowicz, P.
Redlinski, G. Mihaly, and G. Zarand for useful discussions. T.G.R.
acknowledges partial financial support by Brazilian agencies CNPq
(research grant 55.6552/2005-9) and Instituto do Mil\^{e}nio para
Nanoci\^{e}ncia. B.J. was supported by NSF-NIRT award DMR02-10519
and the Alfred P. Sloan Foundation.

\end{document}